\documentclass[10pt,letterpaper]{article}
\usepackage{opex3}

\begin{document}

%% NOTE: TITLE PAGE & TOC NOT USED FOR MANUSCRIPT SUBMISSIONS %%
%\title{Template and style guide for authors submitting to \textit{Optics Express}}
%
%\vskip4pc
%
%\tableofcontents
%\clearpage
%% NO TITLE PAGE FOR OPEX SUBMISSIONS %%

%% START HERE
%%%%%%%%%%%%%%%%%% title page information %%%%%%%%%%%%%%%%%%
\title{Experimental evidence and theoretical modeling of two-photon absorption dynamics in the reduction of intensity noise of solid-state Er:Yb lasers}

\author{Abdelkrim El Amili$^{*}$, Ga\"{e}l Kervella, and M. Alouini}

\address{Institut de Physique de Rennes, Universit\'{e} de Rennes 1, CNRS, Campus de Beaulieu, 35042 Rennes, France}

\email{$^{*}$abdelkrim.elamili@univ-rennes1.fr} %% email address is required



\begin{abstract*} A theoretical and experimental investigation of the intensity noise reduction induced by two-photon absorption in a Er,Yb:Glass laser is reported. The time response of the two-photon absorption mechanism is shown to play an important role on the behavior of the intensity noise spectrum of the laser. A model including an additional rate equation for the two-photon-absorption losses is developed and allows the experimental observations to be predicted.\end{abstract*}

\section{Introduction}
Single-frequency low noise lasers are attractive for a large number of applications such as optical communications, microwave photonics, high resolution spectroscopy, atomic clock interrogation, and metrology. Most of these applications require laser sources providing a narrow linewidth but also an extremely low intensity noise over a wide frequency bandwidth. Although solid-state lasers are known to exhibit narrow spectral widths, they suffer from resonant intensity noise around the Relaxation Oscillation (RO) frequency lying in the range from a few kHz to a few MHz \cite{Alouini01}. RO phenomenon results from the interplay between the population inversion and the intracavity laser intensity. It is present in any class-B laser where the population inversion lifetime is longer than the cavity photon lifetime \cite{McCumber66}. In order to reduce the intensity noise around the RO frequency, several methods have been developed. This can be achieved for instance electronically by a servo-loop acting on the pump intensity \cite{Alouini01,Taccheo96,Zhang03}. However, this method is limited in terms of bandwidth and noise reduction efficiency. Designing a laser which operates in the class-A regime and thus inherently free from relaxation oscillations \cite{Baili07,Feng09} is another solution. Nevertheless, to reach this regime the population inversion lifetime must be shorter than the cavity photon lifetime, making this solution only suitable for lasers including semiconductor active media. Indeed, as far as solid-state lasers are concerned, class-A operation would require kilometric-long cavities. Besides, an interesting alternative for noise reduction consists in inserting a nonlinear absorber inside the laser cavity. While this method was originally proposed to damp down large spiking effects in class-B lasers \cite{Statz65,Pinto67}, it proved to be very effective in reducing the intensity noise around the RO frequency as reported by van Leeuwen \textit{et al.} using two-photon absorption (TPA) in Er,Yb:Glass laser \cite{Leeuwen08}. When inserted into the laser cavity, TPA induces losses which are proportional to the intracavity intensity leading to a significant reduction of the intensity fluctuations. This mechanism can be seen as a feedback loop which instantaneously controls the dynamical behavior of the laser. Indeed, for a given TPA cross-section, increasing the average intensity level leads to a strong suppression of the intensity fluctuations as it is the case when the gain of a feed-back loop is increased. Furthermore, and similarly to a feedback loop, the time response of the TPA mechanism is also an important parameter which must be considered. In particular, the TPA time response has to be much shorter than the time scale of the relaxation oscillations. Our experimental investigations on how TPA affects the laser dynamics and associated noise suggest that the time response of the TPA mechanism must be taken into account. While the model proposed in \cite{Leeuwen08} fails to explain these results, we propose a theoretical approach in which the time dependent TPA mechanism is taken into account by introducing an additional rate equation related to TPA losses. The experimental results are compared to the theoretical predictions and discussed.

\section{Experimental setup and experimental observations}
\begin{figure}[htb]
\centering\includegraphics[width=10cm]{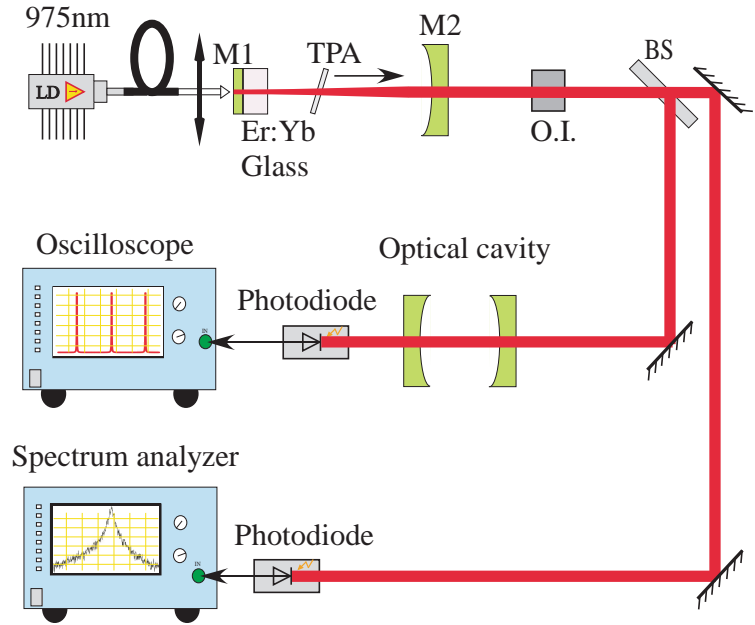}
\caption{Experimental setup. O.I: optical isolator; BS: beam splitter. The arrow shows the direction of displacement of the two-photon absorber in order to change the effective cross-section of the TPA.}\label{Fig1}
\end{figure}
The experimental setup under consideration is depicted in Fig. \ref{Fig1}. The laser used in this experiment is a single-frequency Er,Yb:Glass laser operating at 1560 nm. The laser is formed by a 4.9-cm-long planar-concave cavity. The active medium is a 1.5-mm-long phosphate glass codoped with Erbium and Ytterbium with concentrations of $1.08\times 10^{20}$ ions/cm$^{3}$ and $20\times 10^{20}$  ions/cm$^{3}$ respectively. The first side of the Er,Yb:Glass plate, which acts as the cavity input mirror, is coated for high reflectivity at 1550 nm (R$>99.9\%$) and high transmission (T$=95\%$) at the pump wavelength (975 nm). The cavity is closed by a 5-cm-radius of curvature mirror transmitting $0.5\%$ of the intensity at 1560 nm. A 40-$\mu$m-thick Silicon (Si) plate is inserted into the laser cavity in order to induce the desired intensity dependent losses through two-photon absorption, this plate being perfectly transparent at 1560 nm. Moreover, this Si plate is left uncoated to act as an intracavity \'{e}talon enabling the laser to be single mode. 
The Si plate is mounted on a mechanical stage having three degrees of freedom, namely, one translation and two rotation axes. The rotation axes allow fine adjustment of the plate tilt in order to ensure perfect single mode and linearly polarized oscillation. The translation axis allows the Si plate to be translated along the optical axis of the resonator in order to change the photon density within the Si plate. Indeed, according to the plano-convex architecture of the optical resonator, and to the fact that its length is adjusted close to the limit of stability, the beam diameter on the Si plate increases when the plate is translated towards the output mirror. The active medium is pumped by a cw multimode fiber-coupled laser diode (from Jenoptik) operating at 975 nm. The pump power is fixed at 275 mW during data acquisition. The optical spectrum is continuously analyzed with a Fabry-Perot interferometer to check that the laser remains monomode without any mode hop during data acquisition. The laser intensity is measured using an InGaAs photodiode (Epitaxx Inc, 3.7 MHz bandwidth) and a homemade low noise amplification setup. Finally, the noise spectrum is recorded with an electrical spectrum analyzer (ESA) whose frequency range is 10 Hz-3.6 GHz.\\
\begin{figure}[htb]
\centering\includegraphics[width=10cm]{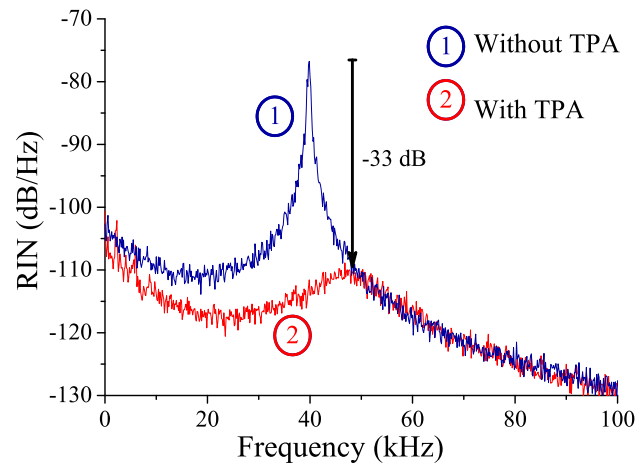}
\caption{RIN spectra of the laser without and with an intracavity two-photon absorber (Si plate). In this example, the insertion of the Si plate leads to a 33 dB reduction of the intensity noise at the relaxation oscillation frequency. ESA resolution bandwidth: 100 Hz; video bandwidth: 100 Hz.}\label{Fig2}
\end{figure}

Fig. \ref{Fig2} reproduces two RIN spectra recorded for two different situations: without (1) and with (2) the Si plate into the laser. In both cases the laser is $\sim$1.2 times above threshold. In accordance with the dynamics of class-B lasers, the RIN spectrum (1) exhibits a strong RO peak at about 40 kHz. By contrast, the RIN spectrum (2) shows that the insertion of the nonlinear absorber close to the active medium, i.e. where the laser photons density is important, leads to 33 dB reduction on the amplitude of the RO peak. Nevertheless, it is worthwhile to notice that the RO frequency is shifted toward high frequencies although the pumping rate is unchanged. Such a frequency shift cannot be predicted by the model proposed in \cite{Leeuwen08}. This behavior in the intensity noise of the laser suggests that the time response of the TPA mechanism must be taken into account as it will be demonstrated in the following.

\section{Theoretical model}
To take into account the time response of the TPA mechanism, we propose to add to the usual rate equations of the laser a new rate equation driving the TPA process. We consider a single-frequency Er,Yb:Glass laser oscillating in cw regime and we assume a homogeneous gain medium described by a quasi-two-level scheme \cite{Laporta93}. The intensity-dependent losses due to the TPA effect are first taken into account in the rate equation driving the laser intensity. However, unlike Ref. \cite{Leeuwen08} these losses being time dependent, they are ruled by a third rate equation as follows:
\begin{eqnarray}
\frac{dN}{dt} & = & -2\sigma IN - \left(w_p + \gamma\right)N + \left(w_p - \gamma\right)N_E, \label{eqn-1}\\
\frac{dI}{dt} & = & \frac{c}{2L} 2e\sigma IN - \gamma_c I - \gamma_{TPA} I, \label{eqn-2}\\
\frac{dP_{TPA}}{dt} & = & \alpha \psi I - \gamma_P P_{TPA}, \label{eqn-3}
\end{eqnarray}
where $\gamma_{TPA} = c P_{TPA}/2L$ is the nonlinear loss rate and $P_{TPA}$ is the time-dependent losses induced by the TPA mechanism. $N$ and $N_E$ are respectively the population inversion and the total Erbium ions (Er$^{3+}$) population, $\gamma$ and $\sigma$ are respectively the population inversion relaxation rate and the emission cross-section of Er$^{3+}$, $w_p$ is the pumping rate, $I$ is the laser photon flux; $e$ and $L$ are respectively the optical lengths of the active medium and of the resonator, $c$ is the velocity of the light. $\gamma_c$ is the photon decay rate which includes both losses of the cold cavity and the slight linear losses of the nonlinear absorber. In the third rate equation which rules the time-dependent loss induced by the TPA mechanism, $\gamma_P/2\pi$ is the inverse of the time response of the TPA mechanism, and $\alpha \psi$ is the effective cross-section of TPA. This effective cross-section takes into account the photon density in the two-photon absorber which is, in general and in particular in our experiment, different from the photon density in the active medium. Thus, $\alpha$ is the ratio between the photon interaction areas in the active medium and in the two-photon absorber. Hence, considering that $w_0$ and $w_{TPA}$ are respectively the beam radius in the active medium and in two-photon absorber $\alpha = w_{0}^2/w_{TPA}^2$.\\

Let us now consider the intensity noise of the laser which is conveniently described by the Relative Intensity Noise (RIN). The RIN level can be determined using a classical approach which is justified by the fact that, in our experiment, the noise level around the RO frequency is higher than the standard quantum noise limit. The parameters which can influence the laser intensity noise, in particular around the RO frequency, are the pump and the cavity loss fluctuations. Moreover, it was shown in \cite{Taccheo98} that the RIN spectrum behavior around the RO frequency can be predicted by computing the intensity transfer function of the laser induced by fluctuations of the cavity losses. This transfer function can be derived from the rate equations by introducing slight perturbations around the equilibrium state of the laser. Hence, small fluctuations of the cavity losses at frequency $\omega$ around the stationary value, i.e., $\gamma_c = \bar{\gamma}_c + \delta \gamma_c \exp(i\omega t)$, induce small fluctuations at the same frequency around the stationary values of the population inversion, i.e.,  $N = \bar{N} + \delta N\exp(i\omega t)$, of the intensity, i.e., $I = \bar{I} + \delta I\exp(i\omega t)$, and of the TPA induced losses, i.e., $P_{TPA} = \bar{P}_{TPA} + \delta P_{TPA}\exp(i\omega t)$. Injecting these expressions in the laser rate equations Eq. \ref{eqn-1}-\ref{eqn-3}, leads to the normalized transfer function:
\begin{eqnarray}
\nonumber H_{\delta \gamma_c}\left(\omega\right) & = & \frac{\delta I/\bar{I}}{\delta \gamma_{c}/\bar{\gamma}_c}\\
 & = & - \frac{\bar{\gamma}_c}{i\omega + \bar{\gamma}_{c}\left(1 - \frac{\bar{N}}{N_{th}}\right) + \gamma_{TPA}\left(1 - \frac{\gamma_{P}}{i\omega + \gamma_P}\right) + \bar{\gamma_c}\frac{\bar{N}}{N_{th}}\frac{2\sigma\bar{I}}{i\omega + 2\sigma\left(\bar{I} + I_{sat}\right)}}, \label{eqn-4}
\end{eqnarray}
where $N_{th} =  \frac{L}{ec\sigma}\bar{\gamma}_c$ is the population inversion at threshold, and $I_{sat} = \frac{w_p + \gamma}{2\sigma}$ is the saturation intensity of the laser. The stationary values of the population inversion, of the intracavity intensity and of the TPA losses read:
\begin{eqnarray}
\bar{N} & = & \frac{1}{2}\left(-\frac{\alpha\psi I_{sat}}{2e\sigma\gamma_{TPA}} + N_{th} + \sqrt{\left(\frac{\alpha\psi I_{sat}}{2e\sigma\gamma_{TPA}} - N_{th}\right)^2 + 4\frac{\alpha\psi I_{sat}}{2e\sigma\gamma_{TPA}}N_0}\right), \label{eqn-5}\\
\bar{I} & = & I_{sat}\left(\frac{N_{0}}{\bar{N}} - 1\right), \label{eqn-6}\\
\bar{P}_{TPA} & = & \frac{\alpha\psi}{\gamma_{P}}\bar{I}, \label{eqn-7}
\end{eqnarray}
where $N_{0} = \frac{w_p+\gamma}{w_p-\gamma}N_E$ is the unsaturated population inversion.\\

Equations Eq. \ref{eqn-5}-\ref{eqn-6} clearly show that the insertion of a two-photon absorber leads to a nonlinear behavior of the population inversion and of the laser intensity. The transfer function in Eq. \ref{eqn-4} predicts two kinds of behaviors depending on the time response of the TPA mechanism for a given TPA cross-section and a given intracavity intensity. Indeed, if the TPA time response is very short as compared to the time scale of the laser relaxation oscillations without TPA, then the two-photon absorber reacts instantaneously to the intensity fluctuations of the laser. Thus, the frequency of the RO peak remains constant whereas its level decreases as the intracavity intensity of the laser increases. By contrast, if the TPA time response becomes close to almost the tenth of the time constant of the laser relaxation oscillations without TPA, then the RO peak decreases as the intracavity intensity increases meanwhile its frequency shifts toward higher frequencies. It must be noted that in this case, the TPA process is less efficient in terms of noise reduction.\\

Our model in which a rate equation for the TPA losses is introduced is able to predict a frequency shift of the RO peak although the pumping rate of the laser is kept constant. In order to validate this model, we compare in the following its predictions with the experimental results.

\section{Experimental validation of the model and discussion}
The experimental setup is described in the first section of the paper. Here, we focus on how the RO peak behaves when the photon density in the Si plate is changed. Rather than changing the intracavity intensity which will be accompanied by a change of the laser parameters, we choose to modify the amount of nonlinear losses by modifying the effective cross-section of the TPA. As described previously, this can be achieved very easily by translating the Si plate along the propagation axis of the laser.\\

Before inserting TPA effects into the laser, we have characterized it by replacing the Si plate by a silica \'{e}talon which introduces almost the same linear losses. In this case the only unknown parameter of the model is the intrinsic intracavity losses, which is adjusted using the experimental intensity-versus-pump characteristic and also by checking that the experimental and theoretical RIN spectra fit together. When the Si plate is inserted back into the laser close to the output mirror, it turns out that the laser threshold remains the same and that the RIN spectra are also the same as compared to that obtained with the silica \'{e}talon. Consequently, the TPA effect can be neglected when the Si plate is close to the output mirror, namely, where the laser beam is sufficiently wide. When the Si plate is moved closer to the active medium, the only unknown parameters are now the effective TPA cross-section and its lifetime.\\
\begin{figure}[htb]
\centering\includegraphics[width=10cm]{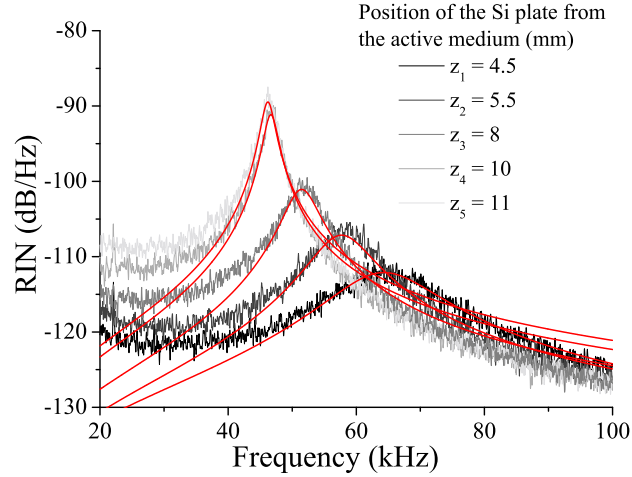}
\caption{RIN spectra of the laser with a Si plate. Each spectrum is recorded for different positions $z$ from the active medium of the TPA plate. The full lines in red are the theoretical transfer functions. The numerical values of the laser parameters used in the calculation are: $N_E = 1.08\times 10^{20}$ cm$^{-3}$; $\gamma = 154$ s$^{-1}$; $e = 0.15$ cm; $L = 5$ cm; $\sigma = 6\times 10^{-21}$ cm$^2$; $\gamma_c = 2.2\times 10^8$ s$^{-1}$; $\alpha(z_1 )\psi = 6.4\times 10^{-22}$ cm$^2$; $\alpha(z_2 )\psi = 2.7\times 10^{-22}$ cm$^2$; $\alpha(z_3)\psi = 1.7\times 10^{-22}$ cm$^2$; $\alpha(z_4 )\psi = 6.9\times 10^{-23}$ cm$^2$; $\alpha(z_5 )\psi = 6.7\times 10^{-23}$ cm$^2$. ESA resolution bandwidth: 100 Hz; video bandwidth: 100 Hz.}
\label{Fig3}
\end{figure}

Fig. \ref{Fig3} reproduces the experimental RIN spectra (in gray color level) recorded for different positions of the Si plate from the active medium, that is, for different photon densities. The theoretical transfer functions are also represented in this figure. The laser parameters that we use in the model are given in the caption Fig. 3. The experimental RIN spectra of Fig. \ref{Fig3}  clearly show that the noise power peak shifts towards higher frequencies meanwhile its level decreases when the Si plate is moved closer to the active medium. As depicted in Fig. \ref{Fig3}, the theoretical model predicts the same behavior, and the theoretical transfer functions match very well the corresponding experimental RIN spectra. Furthermore, we found that the lifetime of the TPA process in our Si plate must be adjusted once at 3 $\mu$s whatever the photon density. This high lifetime value is not so surprising since the carrier's recombination lifetime in silicon semiconductors can be in the range of a few ms to a few ns depending on the crystal purity and whether it is doped or not \cite{Fossum82,Overstraeten87,Alamo87}. These results confirm the fact that the TPA induced losses cannot be considered as instantaneous and that the carrier recombination lifetime must be taken into account when sizing a TPA assisted low noise solid-state laser.\\
\begin{figure}[htb]
\centering\includegraphics[width=10cm]{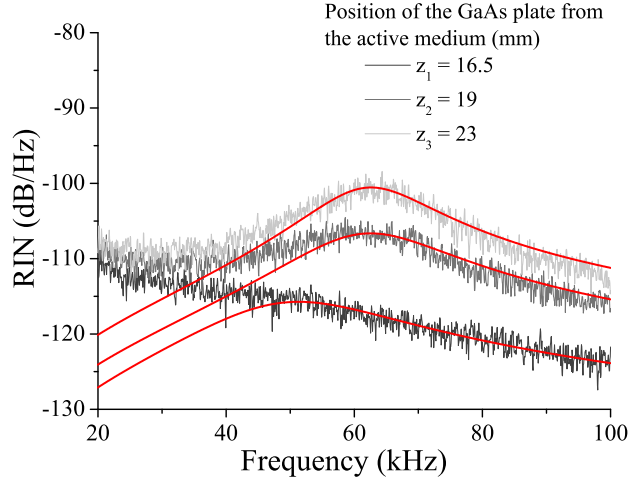}
\caption{RIN spectra of the laser with a GaAs plate. Each spectrum is recorded for different positions z from the active medium of the TPA plate. The full lines in red are the theoretical transfer functions. The numerical values of the laser parameters used are the same as those in Fig. \ref{Fig3} except $\gamma_c = 1.56\times 10^8$  s$^{-1}$; $\alpha(z_1 )\psi = 1.7\times 10^{-19}$ cm$^2$; $\alpha(z_2 )\psi = 7.5\times 10^{-20}$ cm$^2$; $\alpha(z_3 )\psi = 5.5\times 10^{-20}$ cm$^2$. ESA resolution bandwidth: 100 Hz; video bandwidth: 100 Hz.}
\label{Fig4}
\end{figure}

In order to reinforce this conclusion we have replaced the Si plate by a Gallium Arsenide (GaAs) plate cut from a 200-$\mu$m-thick optically polished wafer. Indeed, the recombination lifetime in intrinsic GaAs semiconductor is in the range of few ns, that is much shorter than in Si. In this case, when the photon density increases, our model predicts a strong reduction of the RO peak without any frequency shift. Fig. \ref{Fig4} reproduces both theoretical and experimental RIN spectra obtained with this GaAs plate at different positions from the active medium. As expected, and unlike in the previous experiment, increasing the photon density leads to a significant amplitude reduction the RO peak without any frequency shift. This behavior indicates that the time response of the TPA process in the GaAs plate can be neglected. In other words and in this particular case, the third rate equation that we have introduced can be eliminated adiabatically.

\section{Conclusion}
The intensity noise reduction of Er,Yb:Glass lasers using two-photon absorption is investigated theoretically and experimentally in order to understand the striking behavior observed in the laser intensity noise spectrum. In this respect, we show that the time response of the two-photon absorption mechanism, which is limited by the carrier recombination lifetime, plays an important role. A theoretical model including one additional rate equation that rules the nonlinear losses is derived. It allows recovering the experimental observations. In particular, it predicts that the noise reduction is, in general, accompanied by a shift of the relaxation oscillations even if the pumping rate is kept constant. These theoretical results are confirmed experimentally in and Er,Yb:Glass laser in which a Silicon based plate is inserted. It is shown that increasing the photon density in the Silicon plate, while keeping a constant pumping rate and constant threshold, leads to a reduction of the laser intensity noise, but to an increase of the relaxation oscillations frequency. As a consequence, the noise reduction is less efficient than if the nonlinear absorber would have responded instantaneously to the intracavity intensity fluctuations. This is confirmed by inserting into the laser a GaAs based plate whose recombination lifetime is known to be in the ns range. The good agreement between the theoretical and the experimental results validates the proposed model and makes it well suited to properly design low noise solid-state or semiconductor lasers including two-photon absorbers. This model as well as the related experiments will be extended to the dynamics of dual-frequency lasers including TPA effects where the antiphase noise has also to be reduced.

\section{Acknowledgments}
The authors acknowledge Goulc'hen Loas and Cyril Hamel for technical support. This research was partially funded by the D\'{e}l\'{e}gation G\'{e}n\'{e}rale de l'Armement. Ga\"{e}l Kervella is now with Alcatel-Thales 3-5 lab.

\end{document}